\input epsf

\magnification\magstephalf
\overfullrule 0pt
\openup 1pt
\def\gsim{\raise.3ex\hbox{$\;>$\kern-.75em\lower1ex\hbox{$\sim$}$\;$}}
\font\rfont=cmr10 at 10 true pt
\def\ref#1{$^{\hbox{\rfont {[#1]}}}$}


\font\fourteenbf=cmbx12 scaled\magstep1

\font\tenbfit=cmbxti10
\font\sevenbfit=cmbxti10 at 7pt
\font\fivebfit=cmbxti10 at 5pt
\newfam\bfitfam 
\textfont\bfitfam=\tenbfit  \scriptfont\bfitfam=\sevenbfit
\scriptscriptfont\bfitfam=\fivebfit

\font\eightit=cmti8

\font\tenbfit=cmbxti10
\font\sevenbfit=cmbxti10 at 7pt
\font\fivebfit=cmbxti10 at 5pt
\newfam\bfitfam 
\textfont\bfitfam=\tenbfit  \scriptfont\bfitfam=\sevenbfit
\scriptscriptfont\bfitfam=\fivebfit

\font\tenbit=cmmib10
\newfam\bitfam
\textfont\bitfam=\tenbit%

\font\tenmbf=cmbx10
\font\sevenmbf=cmbx7
\font\fivembf=cmbx5
\newfam\mbffam
\textfont\mbffam=\tenmbf \scriptfont\mbffam=\sevenmbf
\scriptscriptfont\mbffam=\fivembf

\font\tenbsy=cmbsy10
\newfam\bsyfam 
\textfont\bsyfam=\tenbsy%

   
\def\e{\epsilon}

\def\pmb#1{\setbox0=\hbox{#1}
 \kern.05em\copy0\kern-\wd0 \kern-.025em\raise.0433em\box0 }

\def\slash{/\kern-.5em}

\def \half {{\textstyle {1 \over 2}}}

 %


\def\boxit#1{\vbox{\hrule\hbox{\vrule\kern1pt\vbox
{\kern1pt#1\kern1pt}\kern1pt\vrule}\hrule}}

\def\h{\hfill\break}
\parskip=6pt
\parindent=0pt
\hsize=17truecm\hoffset=-5truemm
\vsize=23truecm
\def\footnoterule{\kern-3pt
\hrule width 17truecm \kern 2.6pt}


\catcode`\@=11 

\def\nolabels{\def\wrlabeL##1{}\def\eqlabeL##1{}\def\reflabeL##1{}}
\def\writelabels{\def\wrlabeL##1{\leavevmode\vadjust{\rlap{\smash%
{\line{{\escapechar=` \hfill\rlap{\sevenrm\hskip.03in\string##1}}}}}}}%
\def\eqlabeL##1{{\escapechar-1\rlap{\sevenrm\hskip.05in\string##1}}}%
\def\reflabeL##1{\noexpand\llap{\noexpand\sevenrm\string\string\string##1}}}
\nolabels
\global\newcount\refno \global\refno=1
\newwrite\rfile
\def\defref{$^{{\hbox{\rfont [\the\refno]}}}$\nref}
\def\nref#1{\xdef#1{\the\refno}\writedef{#1\leftbracket#1}%
\ifnum\refno=1\immediate\openout\rfile=refs.tmp\fi
\global\advance\refno by1\chardef\wfile=\rfile\immediate
\write\rfile{\noexpand\item{#1\ }\reflabeL{#1\hskip.31in}\pctsign}\findarg}
\def\findarg#1#{\begingroup\obeylines\newlinechar=`\^^M\pass@rg}
{\obeylines\gdef\pass@rg#1{\writ@line\relax #1^^M\hbox{}^^M}%
\gdef\writ@line#1^^M{\expandafter\toks0\expandafter{\striprel@x #1}%
\edef\next{\the\toks0}\ifx\next\em@rk\let\next=\endgroup\else\ifx\next\empty%
\else\immediate\write\wfile{\the\toks0}\fi\let\next=\writ@line\fi\next\relax}}
\def\striprel@x#1{} \def\em@rk{\hbox{}} 
\def\lref{\begingroup\obeylines\lr@f}
\def\lr@f#1#2{\gdef#1{\defref#1{#2}}\endgroup\unskip}
\newwrite\lfile
{\escapechar-1\xdef\pctsign{\string\%}\xdef\leftbracket{\string\{}
\xdef\rightbracket{\string\}}}

\def\writestop{\def\writestoppt{\immediate\write\lfile{\string\p
ageno%
\the\pageno\string\startrefs\leftbracket\the\refno\rightbracket%
\string\def\string\secsym\leftbracket\secsym\rightbracket%
\string\secno\the\secno\string\meqno\the\meqno}\immediate\closeout\lfile}}
\def\writestoppt{}\def\writedef#1{}
\catcode`\@=12 
\rightline{DAMTP-1999-134}
\rightline{MC/TH-99-13}
\vskip 10truemm
\centerline{\fourteenbf Charm production at HERA}
\vskip 8pt
\centerline{A Donnachie}
\centerline{Department of Physics, Manchester University}
\vskip 5pt
\centerline{P V Landshoff}
\centerline{DAMTP, Cambridge University$^*$}
\footnote{}{$^*$ email addresses: ad@a3.ph.man.ac.uk, \ pvl@damtp.cam.ac.uk}
\bigskip
{\bf Abstract}
The ZEUS data on the charm structure function $F_2^c$ at small $x$
fit well to a single power of~$x$, corresponding to the exchange of
a hard pomeron that is flavour-blind. When combined with the contribution
from the exchange of a soft pomeron, the hard pomeron
gives a good description of elastic
$J/\psi$ photoproduction.

\bigskip\bigskip
We have argued\defref\cudell{
J R Cudell, A Donnachie and P V Landshoff, Physics Letters B448 (1999) 281
} 
that Regge theory should be applicable to the structure function
$F_2(x,Q^2)$ for small $x$ and all values of $Q^2$, however large,
and have shown\defref\twopom{
A Donnachie and P V Landshoff, Physics Letters B437 (1998) 408
} 
that indeed, in its very simplest form, it agrees extremely well with
the available data. In order to fit the data, we introduced a second
pomeron, the hard pomeron, with an intercept a little greater than 1.4; this is
to be contrasted with the soft pomeron that is well-known from soft
hadronic physics, whose intercept is close to 1.08.

Our main message in this paper is that the concept of the hard pomeron,
with an intercept  that is independent of $Q^2$ and 
is a little greater than 1.4,
is supported by the recent ZEUS data\defref\zeuscharm{
ZEUS collaboration: A Breitweg et al, hep-ex/9908012 
}
for the charm structure function $F_2^c$. These data require
only a hard pomeron: the coupling of the soft pomeron to charm is
apparently very small. Hence the data for $F_2^c$
are described by a single power
of $x$. This is shown in figure 1, where the straight lines are
$$
F_2^c(x,Q^2)=f_c(Q^2)x^{-\e _0}
\eqno(1)
$$
with $\e _0=0.44$.
\topinsert
\centerline{\epsfxsize=0.8\hsize\epsfbox{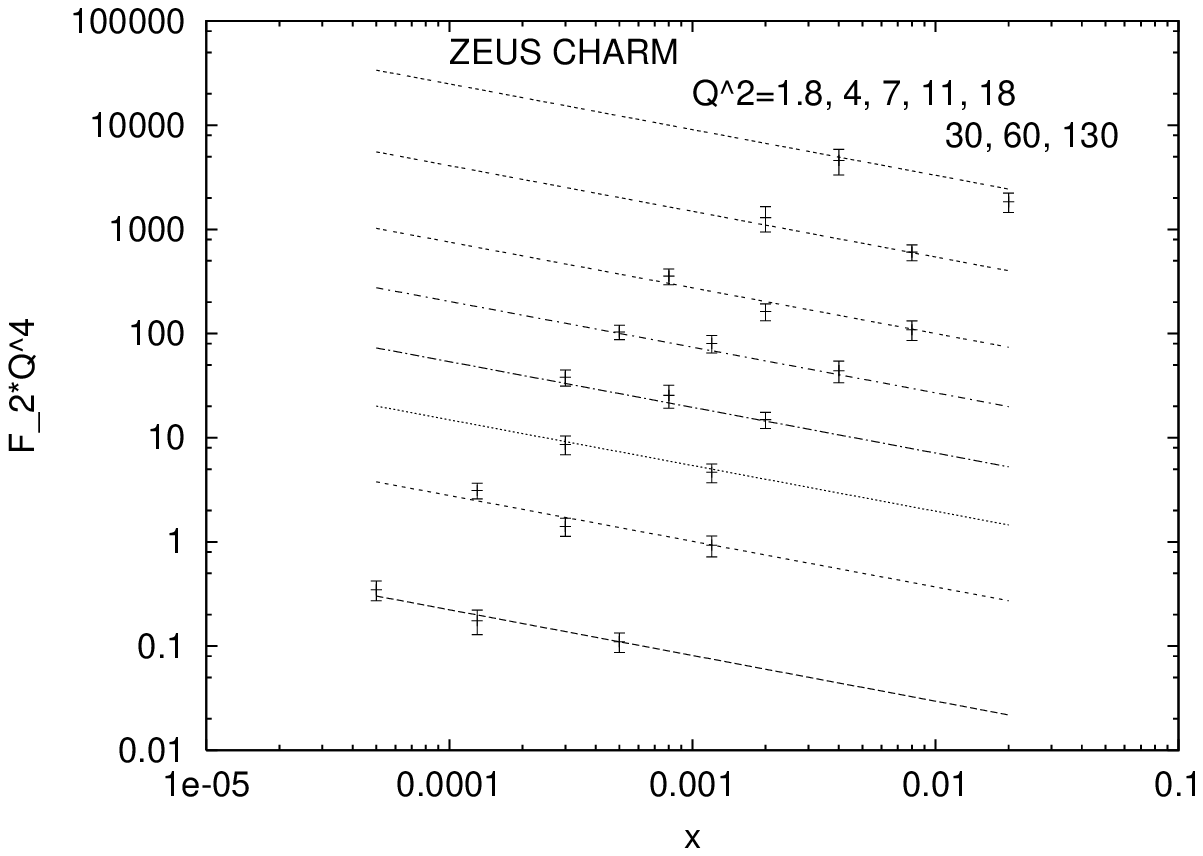}}\h
Figure 1: ZEUS data for $Q^4F_2^c$, fitted to a single fixed power of $x$
\endinsert

In our original fit\ref{\cudell} to the data  for the 
complete structure function
$F_2(x,Q^2)$, we assumed a particular functional form for the coefficient
function $f_0(Q^2)$ that multiplied $x^{-\e _0}$. It had 4 parameters,
and at large $Q^2$ it increased logarithmically with $Q^2$. We have since
found that a form with only 2 parameters works at least as well:
$$
f_0(Q^2)=A_0\left ({Q^2\over Q^2+Q_0^2}\right )^{1+\e _0}
             \left (1+{Q^2\over Q_0^2}\right )^{\half\e _0}
\eqno(2)
$$
With this form, $f_0(Q^2)x^{-\e _0}$ behaves as a $Q^2$-independent
constant times $\nu^{\e _0}$
for large $Q^2$. There is no general theory that explains this behaviour,
though it has been predicted\defref\ermolaev{
B Ermolaev, private communication
}
from the BFKL equation. As we have explained previously\ref{\cudell},
while the large-$Q^2$ behaviour of $f_0(Q^2)$ should surely be calculable
from perturbative QCD, leading-order or next-to-leading-order approximations
are inadequate and at present we do not know how to perform the necessary
all-order resummations.

The fit to $F_2^c$ shown in figure 1 takes
$$
f_c(Q^2)=A_c\left ({Q^2\over Q^2+Q_c^2}\right )^{1+\e _0}
             \left (1+{Q^2\over Q_c^2}\right )^{\half\e _0}
\eqno(3)
$$
In making our fit, we wrote the hard-pomeron contribution to the complete
structure function $F_2(x,Q^2)$ as 
$$
\big (f_0(Q^2)+f_c(Q^2)\big )x^{-\e_0}
$$
with $f_0(Q^2)$ and $f_c(Q^2)$ parametrised as in (2) and (3).
Initially we imposed the constraint that at large $Q^2$ 
the hard pomeron coupling
becomes flavour-blind, so that
$$
A_c Q_c^{-\e _0}={4\over 7}A_0 Q_0^{-\e _0}
\eqno(4)
$$
The factor $4\over 7$ is calculated from squares of quark charges:
${4\over 9}/{({4\over 9}+{1\over 9}+{1\over 9}+{1\over 9})}$.
However, we found that, although it is not excluded that $Q_c^2$ is
somewhat greater than $Q_0^2$, the best fit has $Q_c^2$ close to
$Q_0^2$. That is, the data indicate that the coupling of the hard pomeron
may be flavour-blind even for small $Q^2$. This came as a surprise to us.
Presumably it would imply that the same be true for the proton's
bottom distribution.

With the constraint that $Q_c^2=Q_0^2$, our fit to the ZEUS charm structure
function data, together with nearly 600 data points for $F_2$, corresponding
to $x<0.07$ and $0\leq Q^2\leq 2000$ GeV$^2$, yielded a $\chi^2$ of
less than 1 per data point and
$$
\e _0=0.44~~~~~A_0=0.025~~~~~Q_0^2=8.1\hbox{ GeV}^2
\eqno(5)
$$
More accurate data for $F_2$ are expected soon from HERA, and so the parameter
values will change, as may the tentative conclusion that $Q_c^2=Q_0^2$.

We have already shown\ref{\twopom} that the two-pomeron picture gives a 
good fit to the total cross-section for elastic $J/\psi$ photoproduction,
$\gamma p\to J/\psi\, p$. 
There are now preliminary data \defref\dcs{
H1 Collaboration, submitted to
the International Europhysics Conference on High Energy Physics HEP99, Tampere,
Finland, July 1999
} 
on the differential cross section.
As before\ref{\twopom}, we take the amplitude to be
$$
T(s,t) = i\sum_{i=0,1}\beta_i(t)s^{e_i(t)}e^{-{{1}\over{2}}\pi e_i(t)}
\eqno(6)
$$
We normalise it so that $d\sigma/dt=|T|^2$.
The differential-cross-section data now allow us to make a more informed choice
of the pomeron coupling
functions $\beta_i(t)$. Whereas in elastic $pp$ scattering the
data are in excellent agreement with the hypothesis\defref\elastic{
A Donnachie and P V Landshoff, Nuclear Physics B231 (1983) 189
}
that the soft-pomeron
coupling function is proportional to the square $[F_1(t)]^2$ of the
Dirac electric form factor, the data for $\gamma p\to J/\psi\, p$ rather
need just $F_1(t)$. That is, the proton coupling to the pomeron (either
soft or hard) is proportional to $F_1(t)$, but the pomeron-$\gamma$-$J/\psi$ 
coupling apparently
is flat in $t$. So we use
$$
\beta_i(t)=\beta_{0i}\,F_1(t)~~~~~~~~~~~~~~~~~~~~~~~i=0,1$$$$
F_1(t) = {{4m^2-2.79t}\over{4m^2-t}}{{1}\over{(1-t/0.71)^2}}
\eqno(7)
$$
For the functions $e_i(t)$, which are related to the two pomeron
trajectories by $\alpha_i(t)=1+e_i(t)$, we take
$$
e_0(t) = 0.44+\alpha'_0t~~~~~~~~~~e_1(t) = 0.08+0.25t
\eqno(8)
$$
\topinsert
\centerline{\epsfxsize=0.6\hsize\epsfbox{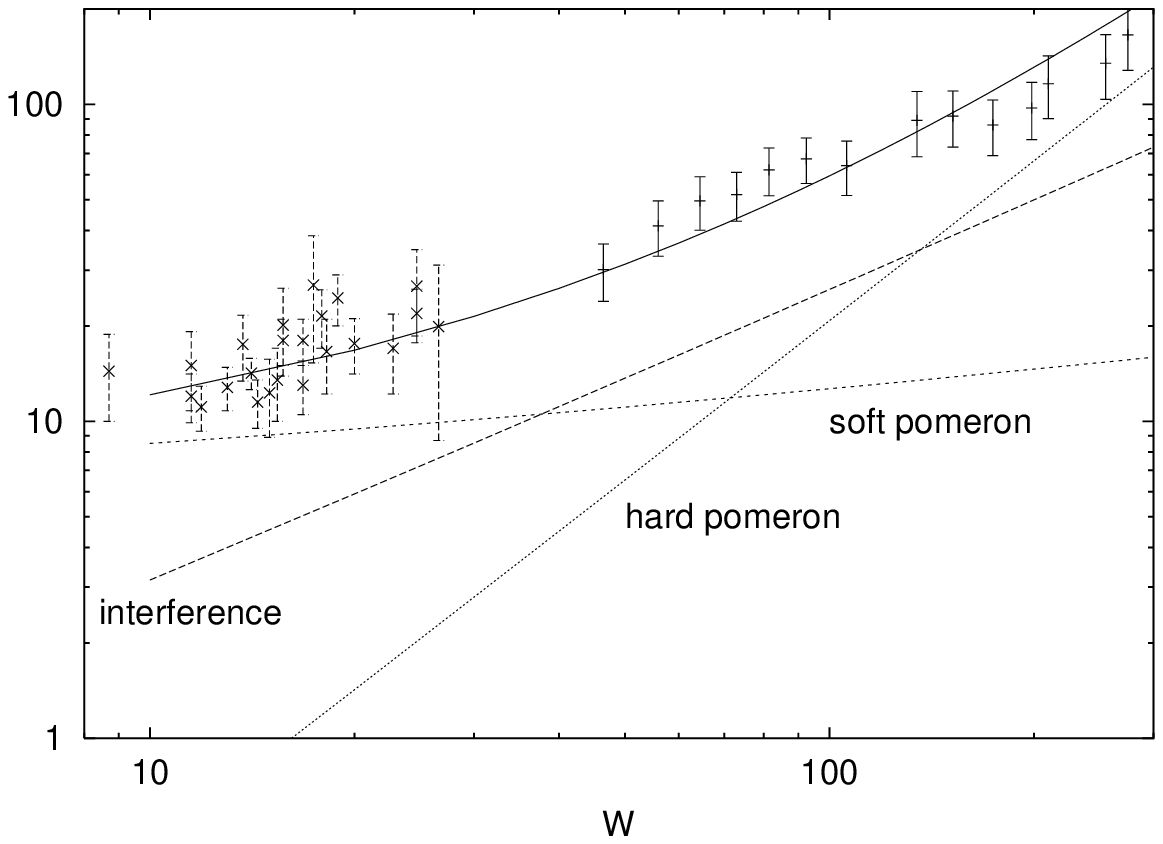}}\h
Figure 2: Fit to the total cross-section for elastic $J/\psi$ 
photoproduction; the data are fixed-target and H1\ref{\dcs}.
The three contributions add up to the solid curve.
\vskip 10mm
\centerline{\epsfxsize=0.455555\hsize\epsfbox{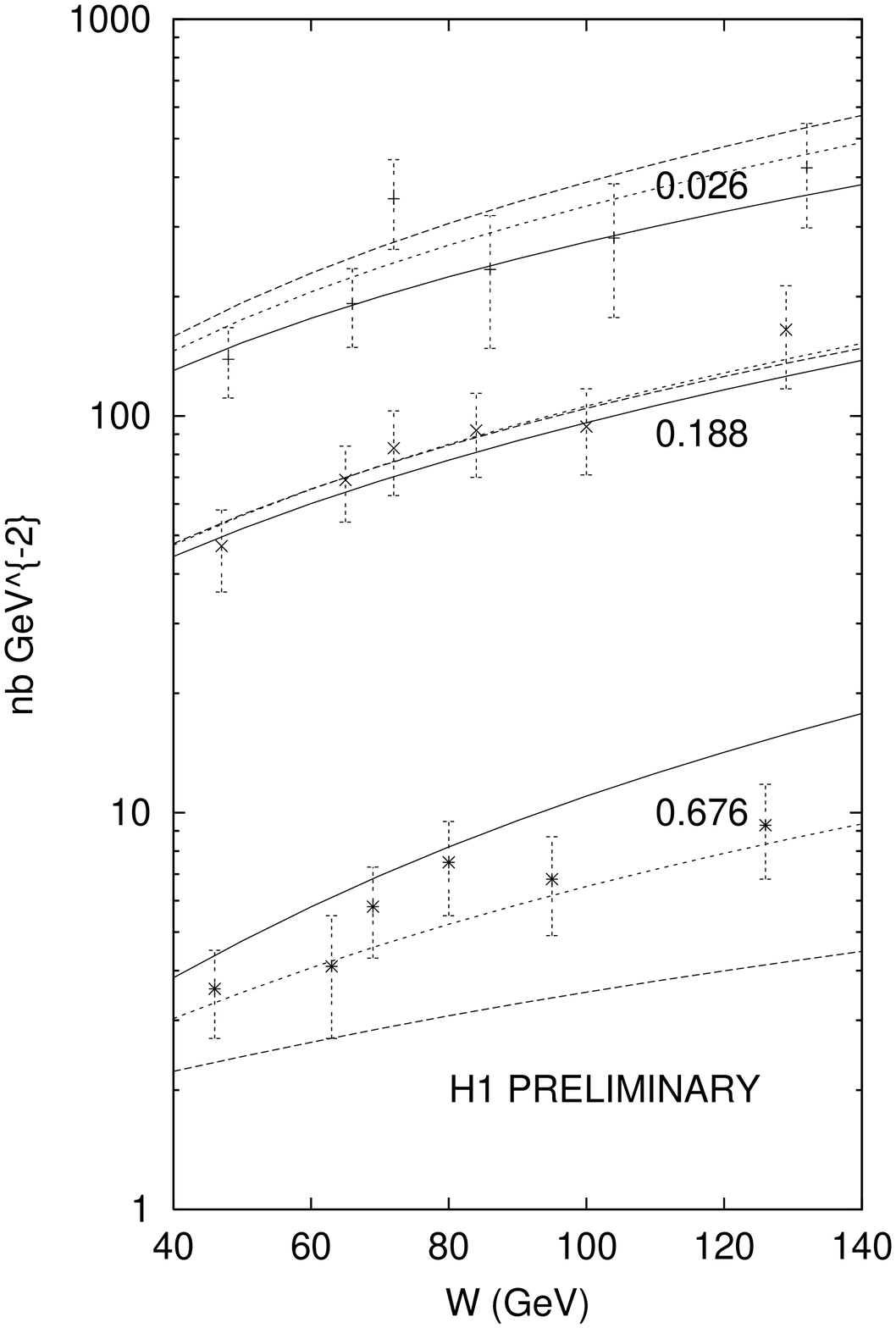}}\h
Figure 3: Fits to the differential cross-section for elastic $J/\psi$ 
photoproduction for three $t$-values and hard pomeron slope
$\alpha'_0=0$ (solid lines), $\alpha'_0=0.1$ (dotted lines)
and $\alpha'_0=0.2$ (dashed lines)

\endinsert

The soft-pomeron trajectory is familiar\ref{\elastic}, but the slope of
the hard-pomeron trajectory is not known. 
The fit shown in figure 2 for the total cross-section is for
$$
\alpha'_0=0.1~~~~~~~~~~\beta _{01}^2=24.6 ~~~~~~~~\beta_{00}=0.038\,\beta _{01}
\eqno(10)
$$
We may obtain almost equally good fits to the total cross section if we make 
different choices of 
$\alpha'_0$, provided we adjust $\beta _{00}$ and $\beta _{01}$:
$$
\alpha'_0=0.0~~~~~~~~~~~\beta _{01}^2=26.4 ~~~~~~~~\beta_{00}=0.028\,\beta _{01}
$$
$$
\alpha'_0=0.2~~~~~~~~~~\beta _{01}^2=23.7 ~~~~~~~~\beta_{00}=0.046\,\beta _{01}
\eqno(11)
$$
Note, though, that $\alpha'_0=0$ strictly is excluded, through $t$-channel 
unitarity\defref\collins{
P D B Collins, {\it Regge theory and high energy physics}, Cambridge
University Press (1977)}.
We show in figure~3 the differential cross-section for these three choices
of $\alpha'_0$. It is evident that a choice somewhere near to 0.1 is a good
one --- though this cannot be a firm conclusion because the data are not
good enough to confirm that (7) is necessarily the correct choice for
$\beta_i(t)$.  However, it is interesting that $\alpha'_0=0.1$
happens to
be the value that is obtained by supposing that the hard pomeron
trajectory is a glueball trajectory, so that there is a 2$^{++}$ glueball
of mass $M$ given by $\alpha_0(M^2)=2$. This corresponds to $M=2370$ MeV,
close to the mass of a 2$^{++}$ glueball candidate reported by
the WA102 collaboration\defref\wa102{
WA102 collaboration: A Barberis et al, Physics Letters B432 (1998) 436
}. (Similarly, there is a 2$^{++}$ glueball candidate at 1930 MeV,
the correct mass for it to lie on the soft pomeron trajectory\defref\wa91{
WA91 collaboration: S Abatzis et al, Physics Letters B324 (1994) 509
}.) The values of 0.0 and 0.2 for $\alpha'_0$ are at the extremes
which the differential cross sections will accept, and limits of $\sim 0.05$
and 0.15 are more reasonable, with of course the above caveat on our choice
of $\beta_i(t)$.

It is not excluded that
there is also a hard-pomeron component present in elastic $\rho$
photoproduction, though there the ratio $\beta _{00}/\beta_{01}$ is
very much smaller. It is possible that the value of $\beta _{00}$
is the same in each case, up to a factor that reflects the different charges
on the active quarks. In either case, $\rho$ or $J/\psi$, 
if the data are parametrised by an effective power rise with energy
$W^{\delta}$, the increase\defref\electro{
ZEUS Collaboration: J. Breitweg et al, Eur Phys J  C6 (1999) 603\h
H1 Collaboration: C  Adloff et al, hep-ex/9902019   
}
of $\delta$ with $Q^2$ may be explained by the
ratio $\beta _{00}/\beta_{01}$ increasing with $Q^2$.

We end with a comment that
the surprisingly complete decoupling of the soft pomeron
in the charm structure function presumably results from
the limited overlap between
the small $c\bar c$ pair and the extended soft pomeron. Justification for
this view is the observation \ref{\twopom} that the soft pomeron contribution
to the proton structure function $F_2$ decreases with increasing $Q^2$ for
$Q^2 \gsim 5$ GeV$^2$. This can be quantified in the dipole-scattering
approach of the Heidelberg model \defref\rueter{M Rueter, Eur.Phys.J. C7 
(1999) 233}, in which an explicit cut-off for the coupling of the soft 
pomeron to small dipoles simulates the phenomenological result of 
[\twopom].
It might then be thought that exactly the same phenomenon would be observed
in $J/\psi$ photoproduction. However the fixed-target data collectively imply 
that there is some contribution at lower energies from the soft pomeron. This
is confirmed by specific fits \ref{\twopom,\rueter} in the two-pomeron 
approach. A resolution of this apparent inconsistency can be obtained by 
postulating that there is an OZI-violating contribution to $J/\psi$ 
photoproduction. Quite apart from the fact that the hadronic decays of the 
$J/\psi$ are by this mechanism, there is clear evidence for an OZI-violating 
contribution to inclusive $J/\psi$ production in hadronic interactions. At 
low energy the $J/\psi$ production cross section from an antiproton beam 
is\defref\corden{M J Corden et al, Physics Letters 68B (1977) 96} is several
times greater than that from a proton beam. This shows that, in $J/\psi$ 
production in hadronic interactions, there is a contribution from the valence 
quarks of the nucleon.
The strength of the coupling of the $J/\psi$ to a light quark-antiquark pair
may be extracted from the production data\defref\green{M B Green, M Jacob
and P V Landshoff, Nuovo Cimento 29A (1975) 123}\defref\gunion{J F Gunion,
Phys.Rev. D12 (1975) 1345}\defref\dlcharm{A Donnachie and P V Landshoff,
Nucl.Phys. B112 (1976) 233}, and is compatible with the hadronic decay rate
of the $J/\psi$. The data on $\Upsilon$ production in hadronic interactions,
in an equivalent region of $x_F$, imply that an OZI-violating mechanism is 
operable there also \defref\dlupsilon{A Donnachie and P V Landshoff, 
Zeits.Phys. C4 (1980) 231}.   
It is not possible to quantify {\it a priori} the OZI-violating contribution
to $J/\psi$ photoproduction as it must arise from complicated $u\bar u$,
$d\bar d$, $s\bar s$ systems.

In conclusion, the fixed power of $x$ found in the ZEUS data for
the charmed structure function is most naturally explained by applying
Regge theory at all $Q^2$. This requires the introduction of a hard pomeron,
just as we have found gives an excellent description of the total
proton structure function $F_2$ and elastic $J/\psi$ photoproduction.

\goodbreak
\bigskip{\eightit
This research is supported in part by the EU Programme
``Training and Mobility of Researchers", Networks
``Hadronic Physics with High Energy Electromagnetic Probes"
(contract FMRX-CT96-0008) and
``Quantum Chromodynamics and the Deep Structure of
Elementary Particles'' (contract FMRX-CT98-0194),
and by PPARC}
\goodbreak
\medskip\immediate\closeout\rfile\writestoppt
\baselineskip=7pt{{\bf References}}\bigskip{\frenchspacing%
\parindent=20pt\escapechar=` \input refs.tmp\bigskip}\nonfrenchspacing
\bye